\documentstyle[12pt]{article}

\begin{document}

\author{N.S.~Ananikian$^{a,b}$\thanks{
e-mail: ananik@jerewan1.yerphi.am}, \and R.G.~Ghulghazaryan$^{b}$\thanks{
e-mail: ghulr@moon.yerphi.am}, \and N.Sh.~Izmailian$^{b,c}$\thanks{
e-mail: izmailan@phys.sinica.edu.tw } \\
$^{a}${\normalsize Department of Physics and Centre for Nonlinear Studies,}\\
{\normalsize Hong Kong Baptist University, Hong Kong, China}\\
$^{b}${\normalsize Department of Theoretical Physics, Yerevan Physics
Institute,}\\
{\normalsize Alikhanian Br.2, 375036 Yerevan, Armenia }\\
$^{c}${\normalsize \ Institute of Physics, Academia Sinica, Taipei 11529, 
Taiwan}}
\title{{\Large {CORRELATION FUNCTIONS OF THE ISING MODEL WITH MULTISITE
INTERACTION ON THE HUSIMI LATTICE}}}
\maketitle

\begin{abstract}
We consider a general spin-1/2 Ising model with multisite interaction on the
Husimi lattice with the coordination number $q$ and derive an analytical
expression of correlation functions for stable fixed points of the
corresponding recurrence relation. We show that for $q=2$ our model
transforms to the two-state vertex model on the Bethe lattice with $q=3$ and
for the case $q=3$, with only nearest neighbour interactions, we transform
our model to the corresponding model on the Bethe lattice with $q=3$ , using
the Yang-Baxter equations.
\end{abstract}

\newpage

\section{ {Introduction}}

Correlation functions play an important role in the understanding of the
critical behavior of the statistical mechanical models. There were received
some exact results for correlation functions and critical exponents for the
Euclidean plane models using conformal invariance \cite{baxter,Cardy}.
Usually Bethe-like (Husimi) lattices considered as embedded in the
infinite-dimensional Euclidean space. However Bethe-like lattices can be
embedded in a two-dimensional space of constant negative curvature (the
hyperbolic plane) with fixed bond angles and lengths. These hierarchical
structures like the Bethe and Husimi lattices can be described as a
conformal (regular) tiling of the hyperbolic plane \cite{m,mi}. It is
believed that several among its interesting thermal properties could persist
for regular lattices, for which the exact calculation is so far intractable.

The exact correlation function for Ising model on the Bethe lattice was 
obtained by Mukamel \cite{mukamel} and Falk \cite{falk} and for
Potts model on the Cayley tree in zero magnetic field by Wang and Wu \cite
{wang}. The two-vertex model on the Bethe lattice with the coordination number 
$q=3$ was explored \cite{Kolesik}, a closed formula for correlation functions 
in stable fixed points of the corresponding recurrence relation was derived and
two types of first-order phase transitions which were distinguished
according to the behavior of correlation functions in the high-temperature
phase were obtained \cite{Kolesik}. Recently Izmailian and Hu derived the 
exact correlation functions for the Ising, BEG and the most general spin -$S$ 
models on the Bethe lattice \cite{izmailian,hu}. Applying combinatorial 
approach and graph expansion Yang and Xu calculated multi-spin correlation 
function of the Ising model on Bethe-type lattice \cite{xu}, they results 
indicates no existence of long range correlation at any finite temperature 
which implies that no finite temperature phase transition occurs. Recently, 
the multisite interaction Ising model on the Husimi lattice was investigated. 
First, it was shown that this approach yields good approximation for the 
ferromagnetic phase diagrams  \cite{a1} and closely matches the exact results 
obtained on a Kagome lattice \cite{a2}. 
The reason for studying the Ising model with multisite interaction is that
it plays an important role in the investigations of real physical systems
such as binary alloys \cite{a3}, classical fluids \cite{a4}, solid $^{3}He$ 
\cite{a5}, liquid bilayers \cite{a6}, and rare gases \cite{a7}.

In this paper we derive a closed formula for correlation functions of the
multisite interaction Ising model on the Husimi lattice for arbitrary
coordination number $q$ and show that our model for the case $q=2$ coincides
with the two-vertex model \cite{Kolesik} and for only nn (nearest neighbour)
interaction Ising model it transforms into the corresponding model on the
Bethe lattice with $q=3$.

The plan of the paper is as follows. In section 2, we define the model,
derive the relevant recurrence relation and magnetization deep within the
tree. In section 3, we derive a closed formula for correlation functions $%
g(n)=\left\langle S_{1}S_{n}\right\rangle -\left\langle S_{1}\right\rangle
\left\langle S_{n}\right\rangle $ , where the symbol $\left\langle \ldots
\right\rangle $ signifies the mean value and $S_{1}$ and $S_{n}$ are spins
on $0$th and \-$\left( n-1\right) $th shells respectively. Section 4 is
devoted to the star-triangle relation between the partition functions of
Ising models on the Husimi and Bethe lattices.

\section{Recurrence relation and magnetization}

The pure Husimi lattice \cite{fractal,sa}, shown in Fig.1, is characterized
by the coordination number $q$, the number of triangles which goes out from
each site, the $0$th-generation is a single central triangle.

\setlength{\unitlength}{1mm}

\begin{center}
\begin{picture}(80,70)
\multiput(27,5)(18,0){2}{\line(1,0){12}}
\multiput(9,65)(36,0){2}{\line(1,0){30}}
\multiput(21,25)(36,0){2}{\line(1,0){6}}
\multiput(9,35)(54,0){2}{\line(1,0){12}}
\multiput(3,55)(72,0){2}{\line(1,0){6}}
\multiput(12,70)(18,0){4}{\line(1,0){6}}
\put(18,20){\line(1,0){48}}
\put(0,50){\line(1,0){84}}

\multiput(0,50)(9,-15){4}{\line(3,-5){3}}
\multiput(3,55)(18,-30){2}{\line(3,-5){15}}
\multiput(9,65)(36,-60){2}{\line(3,-5){3}}
\multiput(30,70)(27,-45){2}{\line(3,-5){6}}
\multiput(45,65)(18,-30){2}{\line(3,-5){3}}
\multiput(66,70)(9,-15){2}{\line(3,-5){6}}
\put(12,70){\line(3,-5){42}}
\put(48,70){\line(3,-5){24}}

\multiput(3,45)(9,15){2}{\line(3,5){6}}
\multiput(18,30)(18,30){2}{\line(3,5){3}}
\multiput(21,15)(27,45){2}{\line(3,5){6}}
\multiput(36,0)(36,60){2}{\line(3,5){3}}
\multiput(48,0)(18,30){2}{\line(3,5){15}}
\multiput(54,0)(9,15){4}{\line(3,5){3}}
\put(12,30){\line(3,5){24}}
\put(30,0){\line(3,5){42}}

\put(28,48){\makebox(0,0)[tl]{$S_{0}^{1}$}}
\put(56,48){\makebox(0,0)[tr]{$S_{0}^{2}$}}
\put(42,26){\makebox(0,0){$S_{0}^{3}$}}
\put(18,64){\makebox(0,0)[tl]{$S_{1}^{1}$}}
\put(30,64){\makebox(0,0)[tr]{$S_{1}^{2}$}}

\end{picture}

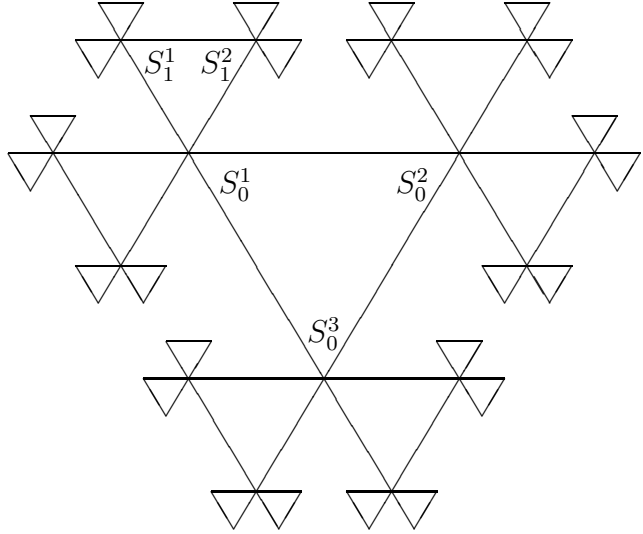
Fig.1 The Husimi lattice with $q=3$ .
\end{center}

The general Hamiltonian with pair, three-site interactions in an external
magnetic field is

\begin{equation}
-\frac{H}{kT}=J_{2} 
\sum_{\left\langle ij\right\rangle }
S_{i}S_{j}+J_{3}\sum_\triangle S_{i}S_{j}S_{k}+h
{\sum }_i S_{i},    \label{1}
\end{equation}
where $S_{i}$ takes values $\pm 1$; the first sum goes over all nearest
neighbour pairs of the lattice sites, the second sum goes over all
triangular faces of the Husimi lattice and the third one goes over all
sites. The advantage of the Husimi lattice introduced here is that for the
models formulated on it, the exact recurrence relation can be derived. From
Fig.1 it is apparent that if the graph is cut at the central triangle, then
it splits up into three identical disconnected pieces and each of them
contains $\gamma $ sub-lattices, where we defined $\gamma =q-1$. Then the
partition function 
\[
Z(N)=\sum_{\{S\}}\exp \left[ -\frac{H}{kT}\right] 
\]
may be written 
\begin{eqnarray}
Z(N) &=&}{\sum_{\left\{ S_{0}\right\}}
\exp \left[
J_{2}(S_{0}^{1}S_{0}^{2}+S_{0}^{1}S_{0}^{3}+S_{0}^{2}S_{0}^{3})+J_{3}S_{0}^{1}S_{0}^{2}S_{0}^{3}+%
\frac{h}{q}(S_{0}^{1}+S_{0}^{2}+S_{0}^{3})\right] \times   \label{2}
\\
&&\times B^{\gamma }(N,S_{0}^{1})B^{\gamma }(N,S_{0}^{2})B^{\gamma
}(N,S_{0}^{3}),  \nonumber
\end{eqnarray}
where $S_{0}^{i}$ are spins of the central triangle and $B(N,S)$ is the
partition function of one sub-lattice containing $N$ shells and with the
root-spin fixed in the state $S$ . Here we use the same notations as in
Kolesik paper \cite{Kolesik}. Because of our Husimi lattice is constructed
from {\it{triangles}} we can introduce statistical weights depending on
three spins located on triangle's vertexes. Introducing the notations

\begin{equation}
w(S_{1},S_{2},S_{3})=\exp \left[
J_{2}(S_{1}S_{2}+S_{1}S_{3}+S_{2}S_{3})+J_{3}S_{1}S_{2}S_{3}+\frac{h}{q}%
(S_{1}+S_{2}+S_{3})\right]    \label{3a}
\end{equation}
and 
\begin{eqnarray}
a_{0} &=&w(-,-,-)=\exp \left[ 3J_{2}-J_{3}-3\frac{h}{q}\right] ,
\label{3b} \\
a_{1} &=&w(+,-,-)=w(-,+,-)=w(-,-,+)=\exp \left[ -J_{2}+J_{3}-\frac{h}{q}%
\right] ,  \label{3c} \\
a_{2} &=&w(+,-,+)=w(+,+,-)=w(-,+,+)=\exp \left[ -J_{2}-J_{3}+\frac{h}{q}%
\right] ,  \label{3d} \\
a_{3} &=&w(+,+,+)=\exp \left[ 3J_{2}+J_{3}+3\frac{h}{q}\right] ,
\label{3e}
\end{eqnarray}
we can present the partition function as follows 
\begin{equation}
Z(N)=\sum_{S_{0}^{1},S_{0}^{2},S_{0}^{3}}
w(S_{0}^{1},S_{0}^{2},S_{0}^{3})B^{\gamma }(N,S_{0}^{1})B^{\gamma
}(N,S_{0}^{2})B^{\gamma }(N,S_{0}^{3}).    \label{4}
\end{equation}
Using the ``cutting apart'' procedure described above, one can derive the
recurrence relation for $B(N,S)$%
\begin{equation}
B(N,S)=\sum_{S_{1},S_{2}}w(S,S_{1},S_{2})B^{\gamma
}(N-1,S_{1})B^{\gamma }(N-1,S_{2}).   \label{5}
\end{equation}
The magnetization of the central site we can express as follows 
\[
m=\left\langle S_{0}^{1}\right\rangle =\left( Z(N)\right) ^{-1}{\sum_{%
S_{0}^{1},S_{0}^{2},S_{0}^{3}} }S_{0}^{1}w\left(
S_{0}^{1},S_{0}^{2},S_{0}^{3}\right) B^{\gamma }(N,S_{0}^{1})B^{\gamma
}(N,S_{0}^{2})B^{\gamma }(N,S_{0}^{3}). 
\]
Let the following variables be introduced 
\begin{equation}
q_{N}(S)=\frac{B(N,S)}{B(N,+)}\mbox{\qquad and\qquad}x_{N}=\frac{B(N,-)}{B(N,+)}.
  \label{7}
\end{equation}
Then using~(\ref{5}) one can write the recurrence relation for $x_{N}$%
\begin{equation}
x_{N}=f\,\left( x_{N-1}\right) ,\quad f(x)=\frac{a_{2}+2a_{1}x^{\gamma
}+a_{0}x^{2\gamma }}{a_{3}+2a_{2}x^{\gamma }+a_{1}x^{2\gamma }}.
\label{8}
\end{equation}
The $x_{N}$ has no direct physical meaning, but through it one can express
the magnetization of the central site 
\begin{equation}
m=\frac{a_{3}+a_{2}x_{N}^{\gamma }-a_{1}x_{N}^{2\gamma }-a_{0}x_{N}^{3\gamma
}}{a_{3}+3a_{2}x_{N}^{\gamma }+3a_{1}x_{N}^{2\gamma }+a_{0}x_{N}^{3\gamma }}
  \label{9}
\end{equation}
and other thermodynamic parameters, since we can say that the $x_{N}$
determine the states of the system.

In this paper we restrict our treatment to ferromagnetic case ( $J_{2}>0$
and $J_{3}>0$) and antiferromagnetic one ($J_{3}<0$ or/and $J_{2}<0$) at
high temperatures when our one dimensional mapping~(\ref{8}) has stable fixed
points only. For the antiferromagnetic case at low temperatures $T<T^{*}$
the single fixed point $x_{0}$ become unstable when 
\[
\frac{\partial }{\partial \,x}\,f(x)\mid _{x=x_{0}}<-1 
\]
and a so-called period doubling bifurcation occurs: the recursive sequence $%
\{x_{N}\}$ converges now not to the single fixed point but to the stable
two-cycle $\{x_{1},x_{2}\}$. This phase should be explained as an arising of
a two-sublattice phase such that $x_{1}$ and $x_{2}$ determine the states on
each sublattice \cite{fractal,sa,a1}. In terms of one dimensional maps the
condition of the second order phase transition point of ferromagnetic models
is 
\[
\frac{\partial }{\partial \,x}\,f(x)\mid _{x=x_{c}}=1 
\]
and the second order phase transition point of antiferromagnetic models \cite
{axe} is 
\[
\frac{\partial }{\partial \,x}\,f(x)\mid _{x=x_{c}}=-1. 
\]
We restrict our treatment to stable fixed points of the
mapping~(\ref{8}) because
the procedure applied later usable for fixed points only (see Sect. 3). The
fixed point $x$ of the recurrence relation~(\ref{8}) is a solution of the equation
\begin{equation}
x=f\,(x) \textstyle{\qquad or\qquad} a_{1}x^{2\gamma +1}-a_{0}x^{2\gamma
}+2a_{2}x^{\gamma +1}-2a_{1}x^{\gamma }+a_{3}x-a_{2}=0.  \label{10}
\end{equation}
Thus, for the stable fixed point $x$ one obtains for the magnetization 
\begin{equation}
m={\lim_{N\rightarrow \infty } }\frac{1-x_{N}^{\gamma }\,x_{N+1}}{%
1+x_{N}^{\gamma }\,x_{N+1}}=\frac{1-x^{q}}{1+x^{q}}.  \label{11}
\end{equation}

\section{Correlation Functions}

One can define correlation functions on the Husimi lattice as follows 
\[
g(n)=\left\langle S_{1}S_{n}\right\rangle -\left\langle S_{1}\right\rangle
\left\langle S_{n}\right\rangle 
\]
where spin $S_{1}$ belongs to the central triangle ($0$th shell) and $S_{n}$
located somewhere in the $(n-1)$th shell. Using
denotation~(\ref{7}) and
recurrence relation~(\ref{5}) we can express $\left\langle S_{1}S_{n}\right\rangle
$ as follows 
\begin{eqnarray}
\left\langle S_{1}S_{n}\right\rangle &=&{%
\lim_{N\rightarrow \infty } }\frac{\sum S_{1}q_{N}^{\gamma
}(S_{1})w(S_{1},y_{1},z_{1})q_{N}^{\gamma }(y_{1})q_{N}^{\gamma
-1}(z_{1})w(z_{1},y_{2},z_{2})q_{N-1}^{\gamma }(y_{2})\cdots }{\sum
q_{N}^{\gamma }(S_{1})w(S_{1},y_{1},z_{1})q_{N}^{\gamma
}(y_{1})q_{N}^{\gamma -1}(z_{1})w(z_{1},y_{2},z_{2})q_{N-1}^{\gamma
}(y_{2})\cdots }  \nonumber \\
&&\frac{\cdots q_{N-n+2}^{\gamma
-1}(z_{n-1})w(z_{n-1},y_{n},S_{n})q_{N-n+1}^{\gamma
}(y_{n})q_{N-n+1}^{\gamma }(S_{n})S_{n}}{\cdots q_{N-n+2}^{\gamma
-1}(z_{n-1})w(z_{n-1},y_{n},S_{n})q_{N-n+1}^{\gamma
}(y_{n})q_{N-n+1}^{\gamma }(S_{n})}.   \label{12}
\end{eqnarray}
Let us define following matrices with elements 
\begin{eqnarray}
M(z_{i},z_{i+1}) &=&{\sum_{y_{i+1}} }q_{N-i+1}^{\gamma
-1}(z_{i})w(z_{i},y_{i+1},z_{i+1})q_{N-i}^{\gamma }(y_{i+1}),
\label{13a} \\
A^{\prime }(S_{n},S_{1}) &=&S_{n}S_{1}q_{N}(S_{1})q_{N-n+1}^{\gamma }(S_{n}),
  \label{13b} \\
A(S_{n},S_{1}) &=&q_{N}(S_{1})q_{N-n+1}^{\gamma }(S_{n}).
\label{13c}
\end{eqnarray}
Note that for stable fixed points of the mapping~(\ref{8}) the difference between $%
q_{N}(S)$ and $q_{N-i}(S)$ disappears for $N\rightarrow \infty $ and for all
finite values $i$ . Hence, for stable fixed points labeling rows (columns)
of these matrices as $+,-$ from left to right (from up to down),
by~(\ref{13a}),~(\ref{13b}) and~(\ref{13c}) we have
\begin{eqnarray}
M &=&\left( 
\begin{array}{cc}
a_{3}+a_{2}x^{\gamma } & a_{2}+a_{1}x^{\gamma } \\ 
x^{\gamma -1}(a_{2}+a_{1}x^{\gamma }) & x^{\gamma -1}(a_{1}+a_{0}x^{\gamma })
\end{array}
\right) ;   \label{14a} \\
A &=&\left( 
\begin{array}{cc}
1 & x \\ 
x^{\gamma } & x^{\gamma +1}
\end{array}
\right) \textstyle{\qquad and\qquad} A^{\prime }=\left(
\begin{array}{cc}
1 & -x \\ 
-x^{\gamma } & x^{\gamma +1}
\end{array}
\right) .    \label{14b,c}
\end{eqnarray}
Using~(\ref{13a}) - (\ref{14b,c}) we can rewrite~(\ref{12}) as follows
\begin{equation}
\left\langle S_{1}S_{n}\right\rangle =\frac{Tr(A^{\prime }M^{n})}{Tr(AM^{n})}%
.   \label{15}
\end{equation}
Using~(\ref{10}) it is easy to show that matrices
$M$~(\ref{14a})
and $A$~(\ref{14b,c})
commute. Thus, they can be diagonalized simultaneously. After performing
this diagonalization we have 
\begin{equation}
\left\langle S_{1}S_{n}\right\rangle =\left( \frac{1-x^{q}}{1+x^{q}}\right)
^{2}+\frac{4x^{q}}{\left( 1+x^{q}\right) ^{2}}\left( \frac{\lambda _{2}(x)}{%
\lambda _{1}(x)}\right) ^{n},
\end{equation}
where 
\begin{eqnarray}
\lambda _{1}(x) &=&a_{3}+2a_{2}x^{\gamma }+a_{1}x^{2\gamma },
\label{17a} \\
\lambda _{2}(x) &=&x^{\gamma -1}\left( a_{0}x^{\gamma }-a_{2}x+a_{1}\left(
1-x^{\gamma +1}\right) \right)   \label{17b}
\end{eqnarray}
are the eigenvalues of matrix $M$. Note that, for stable fixed points $%
\left\langle S_{1}\right\rangle =\left\langle S_{n}\right\rangle =m$. Thus,
using formula~(\ref{11}) for $m$ we can reveal the following formula for
correlation functions 
\begin{equation}
g(n)=\frac{4x^{q}}{\left( 1+x^{q}\right) ^{2}}\left( \frac{\lambda _{2}(x)}{%
\lambda _{1}(x)}\right) ^{n}    \label{18}
\end{equation}
where $\lambda _{1}(x)$ and $\lambda _{2}(x)$ are defined
in~(\ref{17a}),~(\ref{17b}).

One can prove easily that the correlation function decreases exponentially
with the distance $n$ independently of $x$, which is a consequence of the
infinite dimension of the Husimi lattice.

For a period doubling bifurcation of the mapping~(\ref{8}) the matrices
corresponding to the $M$ and $A$ matrices and depending on the $x_{1}$ and $%
x_{2}$ points of the two cycle $\{x_{1},x_{2}\}$ do not commute, so the
procedure applied above is unusable for this case.

\section{Connection with the two-vertex and nn interaction Ising models on
the Bethe lattice}

In the previous section we have derived a closed formula for correlation
functions on the Husimi lattice. Now we shall show that for $q=2$ our model
coincides with the two-vertex model on the Bethe lattice with $q=3$ . First
we recall the definition of the model. Let us consider the Bethe lattice
with $N$ shells and the coordination number $q=3$. The central point $O$ and
the adjacent edges are considered as the $0$th shell. Each edge, including
the free ends on the surface of the tree, can be in one of the two distinct
spin states $S\in \left\{ +,-\right\} $. To each vertex one ascribe the
statistical weight $w(S_{1},S_{2},S_{3})$ depending on the number of
incident edges in the $(+)$-state. Let $a_{i}$ denote the corresponding
statistical weight of the vertex with $i$ incident edges in the state $+$.
Using notation~(\ref{3a}) and Fig. 2 it is easy to see now that our models coincide.

\setlength{\unitlength}{1mm}

\begin{center}
\begin{picture}(80,70)

\put(40,12.2){\line(0,1){25.6}}
\put(16,52.2){\line(0,1){12.8}}
\put(64,52.2){\line(0,1){12.8}}
\put(28,5){\line(5,3){12}}
\put(40,37.8){\line(5,3){24}}
\put(4,45){\line(5,3){12}}
\put(52,5){\line(-5,3){12}}
\put(40,37.8){\line(-5,3){24}}
\put(76,45){\line(-5,3){12}}

\thicklines
\put(28,5){\line(1,0){24}}
\put(4,45){\line(1,0){72}}
\put(4,45){\line(3,5){12}}
\put(28,5){\line(3,5){36}}
\put(16,65){\line(3,-5){36}}
\put(64,65){\line(3,-5){12}}

\put(30,47){\makebox(0,0)[bl]{$S_{1}$}}
\put(50,47){\makebox(0,0)[br]{$S_{2}$}}
\put(44,23){\makebox(0,0)[bl]{$S_{3}$}}
\put(40,40.8){\makebox(0,0){$O$}}

\end{picture}

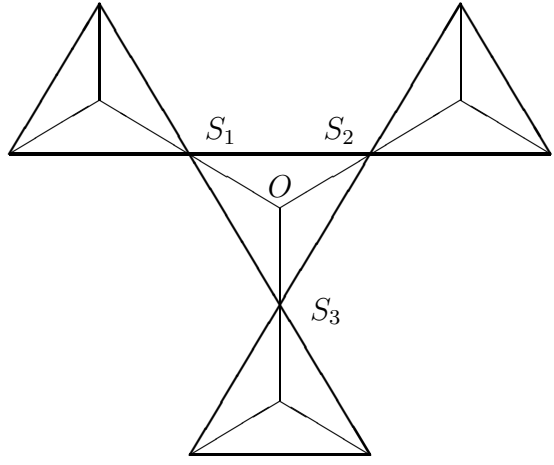
Fig.2 The Husimi and Bethe lattices with $q=2$ and $q=3$ respectively.
\end{center}

In the following we shall show that our model with only nn interaction ($%
J_{3}=h=0$) and $q=3$ transforms to the nn interaction Ising model on the
Bethe lattice with $q=3$. In this case we shall use widely known star-triangle
transformation (or the Yang -Baxter equations) \cite{baxter} to transfer our
model to the Bethe lattice one. The star-triangle transformation replaces a
star consisting of three spins interacting with a central spin, by a
triangle of spins interacting with one another and vice versa (Fig.3).

\setlength{\unitlength}{1mm}

\begin{center}
\begin{picture}(80,30)

\put(6,5){\line(5,3){12}}
\put(30,5){\line(-5,3){12}}
\put(18,12.2){\line(0,1){12.8}}

\put(50,5){\line(1,0){24}}
\put(50,5){\line(3,5){12}}
\put(74,5){\line(-3,5){12}}

\put(40,15){\vector(1,0){5}}
\put(40,15){\vector(-1,0){5}}

\put(20,14){\makebox(0,0)[bl]{$L$}}
\put(54,14){\makebox(0,0)[br]{$J_{2}$}}
\put(19,25){\makebox(0,0)[bl]{$S_{1}$}}
\put(31,5){\makebox(0,0)[tl]{$S_{2}$}}
\put(5,5){\makebox(0,0)[tr]{$S_{3}$}}
\put(18,10){\makebox(0,0)[t]{$S_{0}$}}
\put(63,25){\makebox(0,0)[bl]{$S_{1}$}}
\put(75,5){\makebox(0,0)[tl]{$S_{2}$}}
\put(49,5){\makebox(0,0)[tr]{$S_{3}$}}
\end{picture}

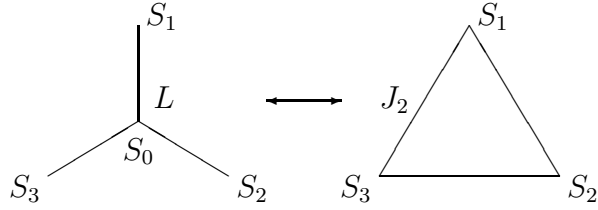
Fig.3 The star-triangle transformation.
\end{center}

Let us consider the simplest case in which the interactions of all
directions are equal. Then summing over the central spin in the partition
function of a star and equaling it to the partition function of a triangle
with some coefficient $R$, we have 
\[
{\sum_{S_{0}} }\exp L\left(
S_{0}S_{1}+S_{0}S_{2}+S_{0}S_{3}\right) =R\,\exp J_{2}\left(
S_{1}S_{2}+S_{1}S_{3}+S_{2}S_{3}\right) 
\]
where $L$ and $J_{2}$ are the pair interaction parameters of the Bethe and
Husimi lattices respectively. After performing the summation over $S_{0}$
and substituting $S_{i}=\pm 1$, $i=1,2,3$ we obtain the Yang-Baxter
equations in the form 
\begin{eqnarray}
2\cosh (3L) &=&R\,\exp (3J_{2}),   \label{19.a} \\
2\cosh (L) &=&R\,\exp (-J_{2}),    \label{19.b}
\end{eqnarray}
From notation~(\ref{3a}) it's follows
\begin{equation}
a_{0}=a_{3}=\exp (3J_{2})\textstyle{\qquad and\qquad}a_{1}=a_{2}=\exp (-J_{2}).
\label{20ab}
\end{equation}
Due to summation over spins located inside of triangles in star-triangular
transformation one can obtain 
\[
g_{H}(n)=g_{B}(2n), 
\]
where $g_{H}(n)$ and $g_{B}(2n)$ are the correlation functions of the Husimi
and Bethe lattices respectively. Thus, for correlation functions of the nn
interaction Ising model on the Bethe lattice we have 
\[
g_{B}(n)=\frac{4x^{3}}{(1+x^{3})^{2}}r^{n}(x) 
\]
where 
\[
r(x)=\frac{x(\exp (2L)-x)}{x^{2}+\exp (2L)}. 
\]
Here we used the fact that $x$ is a stable fixed point of the corresponding
recurrence relation on the Bethe lattice and it coincides with Husimi's
lattice one. Here also one can prove that, independently of $x$, the
correlation function decreases exponentially with the distance $n$, which is
a consequence of the infinite dimension of the Bethe lattice.

From~(\ref{19.a}),~(\ref{19.b}) after simple algebra one can obtain
\begin{equation}
\sinh 2L\sinh 2J_{2}=k^{-1}    \label{21.a}
\end{equation}
where 
\begin{equation}
k^{-1}=\frac{\sinh ^{3}2L}{2\left( \cosh 3L\cosh ^{3}L\right) ^{\frac{1}{2}}}%
.    \label{21.b}
\end{equation}
First the star-triangle transformation was used for triangle and honeycomb
lattices \cite{baxter}. It was obtained that for triangle and honeycomb
lattices the condition of the critical point is $k=1$. For the ferromagnetic
Ising model on the Bethe lattice it is well known that the critical point is 
\[
L_{c}=\frac{1}{2}\ln \frac{q}{q-2} 
\]
where $q$ is the coordination number of the lattice.
From~(\ref{21.b}) follows
that for the case $q=3$ parameter $k$ is unequal to $1$ ($k=\frac{3\sqrt{14}%
}{8}$). From~(\ref{21.a}) one can obtain the critical point of the ferromagnetic
Ising model on the Husimi lattice $(J_{2})_{c}=\frac{1}{4}\ln \frac{7}{3}$
(the last result can be derived in a straightforward manner, see Sect. 2).

\section{Concluding remarks}

In this paper we have derived a closed formula for correlation functions of
the multisite interaction Ising model in the presence of an external
magnetic field. We showed that the correlation function decreases
exponentially with the distance which indicates no existence of long range
correlation at any finite temperature this is a consequence of the infinite
dimension of the Husimi lattice in Euclidean space. This result is in good
agreement with the results obtained in papers \cite{Kolesik,xu}. All our
obtained formulae contain only statistical weights $a_{i}$, which depend on
the spins located in the vertexes of triangles. Hence, our formulae can be
used also for other types of Hamiltonians where nn (nearest neighbour) and
three-site interactions are present, for example Potts model, etc.

The star-triangle relation (the Yang-Baxter equations) maps a
high-temperature model on the Bethe lattice with $q=3$ to the
high-temperature model on the Husimi cactus. Together with nonanalytic
behavior of the recurrence relations of the Bethe and Husimi lattices one
can use them to obtain the critical exponents and correlation functions too.
In conclusion, let us note that the results deduced on the Husimi and
Bethe-like lattices are in good agreement with results obtained on planar
lattices, and according to Gujrati, Bethe or Bethe-like (Husimi cactus)
calculations are more reliable than conventional mean-field calculations 
\cite{Gujrati}.

We are grateful to Professors A. Belavin, R. Flume, B. Hu and M. Roger for
useful discussions.  We would like to thank Professor J. Monroe for calling
our attention to Ref. \cite{falk}. 

This work was partly supported by the Grant INTAS-96-690 and the grants from
the Hong Kong Research Grants Council and the Hong Kong Baptist University.
One of us (N. Sh. I.) thanks the National Science Council of the Republic of
China (Taiwan) for financial support under grant No. NSC 87-2112-M-001-046. 

\pagebreak


\begin{thebibliography}{99}
\bibitem{baxter}  R. J. Baxter,\it{\ Exactly Solved Models in
Statistical Mechanics} (Academic Press, London, 1981).

\bibitem{Cardy}  J. L. Cardy, In \it{''Phase Transitions and Critical
Phenomena''}, Vol. 11 (C. Domb and J. L. Lebowitz), Ch. 2. (Academic Press,
London, 1987).

\bibitem{Kolesik}  M. Kolesik, \it{Int. J. Mod. Phys. } {B6, }3469
(1992).

\bibitem{izmailian}  Chin-Kun Hu and N. Sh. Izmailian, \it{Phys. Rev}.
 {E 58} (1998), in press.

\bibitem{hu}  N. Sh. Izmailian and Chin-Kun Hu,\it{\ Physica } {A 254}%
, 198 (1998).

\bibitem{mukamel}  D. Mukamel, \it{Phys. Lett.} A50, 339 (1974).

\bibitem{falk}  H. Falk,\it{\ Phys. Rev.}  {B12}, 5184 (1974).

\bibitem{wang}  Y. K. Wang, F. Y. Wu,\it{\ J. Phys.} {\ A9}, 593
(1976).

\bibitem{xu}  Z.R. Yang, Chang Ye Xu, \it{Commun. Theor. Phys.} Vol.
 {22}, 419-424 (1994).

\bibitem{m}  R. Mosseri and J. F. Sadoc,\it{\ J. Phys. France Lett}.%
 {\ 43}, L249 (1982).

\bibitem{mi}  J. A. de Miranda-Neto and Fernando Moraes, \it{J. Phys. I
France}  {3,} 29 (1993).

\bibitem{fractal}  N. S. Ananikian et. al, \it{Fractals, }Vol. {\ 5%
}, 175 (1997).

\bibitem{sa}  N. S. Ananikian and S.K.Dallakian, Physica  {D107}, 75
(1997).

\bibitem{a1}  J. L. Monroe, \it{J. Stat. Physics, } {65, }255
(1991).

\bibitem{axe}  N. S. Ananikian, A. Z. Akheyan, \it{Phys. Lett.}  {%
A 186}, 171 (1994).

\bibitem{a2}  X. N. Wu and F. Y. Wu, \it{J. Phys.} {\ A}\it{\
: Math. Gen. } {22}, L1031 (1989).

\bibitem{a3}  D. F. Styer, M. K. Phani and J. I. Lebowitz, \it{Phys. Rev.%
}  {B34}, 3361 (1986).

\bibitem{a4}  M. Grimsditch, P. Loubeyre and A. Polian, \it{Phys. Rev.}
 {B33}, 7192 (1986).

\bibitem{a5}  M. Roger, J. H. Hetherington and J. M. Delrieu, \it{Rev.
Mod. Phys.}  {55}, 1 (1983).

\bibitem{a6}  H. L. Scott, \it{Phys. Rev.}  {A37}, 263 (1988).

\bibitem{a7}  J. A. Barker, \it{Phys. Rev. Lett.}  {57}, 230
(1986).

\bibitem{Gujrati}  P. D. Gujrati, \it{Phys. Rev. Lett.}  {74}, 809
(1995).
\end{thebibliography}
\end{document}